\documentclass{emulateapj}

\usepackage{ulem}
\usepackage{longtable}
\usepackage{graphicx}
\usepackage{amssymb,amsmath}
\usepackage{grffile}

\shorttitle{Evidence for two-component jet in Sw J1644+57}

\begin{document}
\title{Quasi-Periodic Variations in X-ray Emission and Long-Term Radio Observations: Evidence for a Two-Component Jet in Sw J1644+57}
\author{Jiu-Zhou Wang\altaffilmark{1}, Wei-Hua Lei\altaffilmark{1,5,6}, Ding-Xiong Wang\altaffilmark{1},  Yuan-Chuan Zou\altaffilmark{1}, Bing Zhang\altaffilmark{3,4}, He Gao\altaffilmark{3} and Chang-Yin Huang\altaffilmark{1,2}}
\altaffiltext{1}{School of Physics, Huazhong University of Science and Technology, Wuhan, 430074, China.\\ Email: leiwh@hust.edu.cn; dxwang@hust.edu.cn;  zouyc@hust.edu.cn}
\altaffiltext{2}{School of Mathematics and Statistics, Huazhong University of Science and Technology, 430074, Wuhan, China}
\altaffiltext{3}{Department of Physics and Astronomy, University of Nevada Las Vegas, 4505 Maryland Parkway, Box 454002, Las Vegas, NV 89154-4002, USA;  zhang@physics.unlv.edu}
\altaffiltext{4} {Kavli Institute for Astronomy and Astrophysics and Department of Astronomy, Peking University, Beijing 100871, China}
\altaffiltext{5} {Key Laboratory for the Structure and Evolution of Celestial Objects, Chinese Academy of Sciences, Kunming 650011, China}
\altaffiltext{6} {Purple Mountain Observatory, Chinese Academy of Sciences, Nanjing 210008, China}

\begin{abstract}
The continued observations of Sw J1644+57 in X-ray and radio bands accumulated a rich data set to study the relativistic jet launched in this tidal disruption event. The X-ray light curve of Sw J1644+57 from 5-30 days presents two kinds of quasi-periodic variations: a 200 second quasi-periodic oscillation (QPO) and a 2.7-day quasi-periodic variation. The latter has been interpreted by a precessing jet launched near the Bardeen-Petterson radius of a warped disk. Here we suggest that the $\sim$ 200s QPO could be associated with a second, narrower jet sweeping the observer line-of-sight periodically, which is launched from a spinning black hole in the misaligned direction with respect to the black hole's angular momentum. In addition, we show that this two-component jet model can interpret the radio light curve of the event, especially the re-brightening feature starting $\sim 100$ days after the trigger. From the data we infer that inner jet may have a Lorentz factor of $\Gamma_{\rm j} \sim 5.5$ and a kinetic energy of $E_{\rm k,iso} \sim 3.0 \times 10^{52} {\rm erg}$, while the outer jet may have a Lorentz factor of $\Gamma_{\rm j} \sim 2.5$ and a kinetic energy of $E_{\rm k,iso} \sim 3.0 \times 10^{53} {\rm erg}$.

\end{abstract}

\section{Introduction}
The two-component jet model has been frequently referred to to interpret data of active galactic nuclei (AGNs) and gamma-ray bursts (GRBs). \citet{cccg00} suggested a two-component jet model for BL Lacs and FRI radio galaxies, in which the observational differences of these two kinds of objects just result from the different orientations of the observer. The limb-brightened morphology shown in several radio galaxies can be regarded as the evidence of a slower sheath-jet surrounding a faster spine-jet (e.g., \citealt{ggf04}).  The afterglow data of GRB 030329 also demanded a two-component jet (\citealt{bkp03}):  a narrow ultra-relativistic component responsible for the $\gamma$-ray and early afterglow, and a wide, mildly relativistic component responsible for the radio and optical afterglow after 1.5 days. The rebrightening of XRF 030723 and the chromatic behavior of the broadband afterglow of GRB 080319B have been considered as further evidence for a two-component jet in GRBs (\citealt{hwd04, rea08}). \citet{el99} suggested a two-component jet model with a baryon-poor jet existing within a baryon-rich outflow. The baryon-poor jet may be driven by the Blandford-Znajek (B-Z) mechanism (\citealt{bz77}; \citealt{mt82}; \citealt{wc08}; \citealt{lzl13}), in which the rotational energy of a black hole (hereafter BH) is extracted to power the jet in the form of a Poynting flux via the open field lines penetrating the event horizon. \citet{m03} discussed the possibility of using the coexistence of B-Z and Blandford-Payne (B-P) (\citealt{bp82}) processes as an interpretation of the two-component jets for quasars and microquasars. In the B-P process, a baryon-rich outflow can be launched centrifugally via the open magnetic field lines threading through the disk. It is argued that the baryon-rich jet can also play an important role in the collimation of the central jet (\citealt{el99,t10}). In the GRB context, \cite{pkg05} proposed a two-component jet scenario invoking a wide neutron jet and a narrow proton jet, and calculated the afterglow behavior in detail. \citet{xlz12} proposed a two-component jet model for both GRBs and AGNs, in which the inner and outer jets are powered by the B-Z and B-P processes, respectively. \citet{wdhl05} discussed the polarization of GRB afterglows from two-component jets.

Recently, much attention has been paid on the discovery of the hard X-ray transient event
Swift J16449.3+573451, (``Sw J1644+57'' hereafter, \citealt{bkg11}). This event has been
interpreted as a tidal disruption event (TDE) with jet (\citealt{bgm11, ltc11, bkg11, zbs11}).
Its rich observations in $\gamma$-ray, X-ray, radio, mm, infrared bands provide us a good opportunity to study the underlying physics of launching a relativistic jet/outflow from TDE events. Detailed data also allow us to diagnose the composition and the structure of the jet.

The unusual features of Sw J1644+57 in its super-Eddington X-ray luminosity (\citealt{bkg11}),
bright radio afterglow (\citealt{zbs11}), and a historical stringent X-ray flux upper
limit suggest that this TDE is closely related to the onset of a relativistic jet from a
supermassive BH. The jet is expected to be magnetically dominated (\citealt{bkg11, szfw11, g12}). \citet{lz11} suggested that B-Z process is the plausible mechanism to launch the relativistic jet from this source (\citealt{bz77,mt82, lwm05, lwzz08}),
and they used the available data to constrain the BH spin for Sw J1644+57. It is found
that the BH of this source carries a moderate to high spin, suggesting that BH spin is likely the
crucial factor of powering the jet/outflow from this BH system. In addition, \citet[hereafter LZG13]{lzg13}
interpreted a 2.7-day quasi-periodic variation with noticeable narrow dips in the X-ray light curve
by invoking a precessing, B-Z powered jet collimated by a wind launched from a twisted and warped disk.
Such a picture is reasonable, since very likely the initial orbital plan of the disrupted star
is likely mis-aligned with the equatorial plane of the BH.

Recently, \citet{rmr12} detected a $\sim$ 200s Quasi-Periodic Oscillation (QPO)
from the 2-to-10-keV power spectra of both the Suzaku and XMM-Newton observations of
this source at redshift $z = 0.3534$. It is interesting to note that both quasi-periodic variations
(2.7-days and 200s) were detected a few days after the BAT trigger despite of using different instruments
(\citealt{bkg11, sswk12, rmr12}).

A number of models have been proposed to
interpret the 200s QPO. \citet[hereafter AL12]{al12} regarded this observed QPO
as one of ``3:2 twin peak QPO'', assuming that the second frequency was not observed based on the
resonance in two eigen-modes of disk oscillations (\citealt{ak01}). However, it is not clear
how a disk-origin QPO can be manifested in the signal of jet emission.  Very recently,
\citet[hereafter TMGK14]{tmgk14} assumed the presence of a strong magnetic flux threading the BH, and
suggested a five-stage jet activity to interpret the evolution and radiation characteristics of this
TDE. The 200s QPO is explained based on the jet-disk QPO mechanism proposed by
\citet{mtb12}, and the BH mass and spin are constrained in TMGK14 by considering three scenarios:
(i) a complete or (ii) partial tidal disruption of a lower mass main-sequence star by a supermassive BH;
or (iii) a complete disruption of a white dwarf by an intermediate-mass BH. However, they did not interpret the 2.7-day variation. Here we propose a structured jet model to interpret the two quasi-periodic variations
of Swift J1644+57 based on \citet{lz11} and LZG13. We suggest that the jet consists of an inner and
an outer component. The 200s QPO is related to the inner jet launched from the horizon of a spinning BH
surrounded by a warped accretion disk, and the 2.7-days quasi-periodic variation is associated with
the outer processing jet launched near the Bardeen-Petterson radius.

Besides the two quasi-periodic variations discovered in the X-ray emission, the radio observation extending to about 600 days also revealed unexpected features: about 1 month later, the radio emission showed a surprising re-brightening feature. The late time X-ray flux, however, did not show this re-brightening but instead showed a dramatic transition starting from $\sim 500$ days with a sharp decline. Therefore the X-ray and radio emissions likely have distinct physical origins: the X-rays may originate from internal dissipation within the jet (\citealt{wc12, zwc13}), while the radio emission may originate from the forward shock that propagates into the surrounding material (\citealt{mgm12, cw12}). \citet{bzp12} and \citet{zbm13} suggested that the radio re-brightening is a result of late-time energy injection from the central engine. Based on such a single jet model, \citet{cw12} argued that the outflow gradually transits from a conical jet to a cylindrical one at later times. However, to explain the late-time radio re-brightening, these models require that the energy of the source increases by a factor of 10-20 from 5 to 200 days, and there is no indication for this additional energy injection from X-ray observations. More recently, \citet{kbbp13} argued that the effect of inverse-Compton cooling of electrons in the external shock region by the internal emission (X-rays) could give rise to a flat radio light curve.
Realizing that the early-time X-ray and radio emission must have a separate origin, \citet{lpl12} proposed a two-component jet model. In their model, the inner jet with a high Lorentz factor accounts for the early X-ray emission and the late radio re-brightening, while the outer, slower jet is responsible for the early radio emission. This model can explain radio emission up to 216 days. Our two-component jet model is different: besides interpret the two quasi-period variation features in X-rays, we invoke the inner (faster) jet to interpret the early radio emission, and the outer (slower) jet to interpret the late radio observations. Our model can account for the radio light curve all the way to $\sim$ 600 days.

This paper is organized as follows. In Sect. 2, we present our model to interpret the two quasi-periodic variations in the X-ray flux.  In Sect. 3, we apply the model to the interpret the radio data. We summarize the results in Sect. 4 with some discussion.

\section{Two-Component Jet Model for the 200s and 2.7-day Quasi-Periodic Variations}

It is noted that two quasi-periodic variations of Swift J1644+57 were detected after a few
days since the BAT trigger (\citealt{rmr12, bkg11, sswk12}).
The first is a rough 2.7-day periodicity in the dips of the XRT light curve of Sw J1644+57. This quasi-periodic signal was first pointed out by \citet{bkg11}, more carefully studied by \citet{sswk12}, and then further confirmed by LZG13 using the stepwise filter correlation (SFC) method. The details of the SFC method and its application to GRB light curves are presented in \citet{gzz12}. The second QPO feature was discovered by \citet{rmr12}, who produced the light curves of Sw J1644+57 over the 0.2-to-10.0-keV energy band using all the observational data from different detectors. They Fourier transformed these light curves to obtain their power density spectra. The 2-to-10-keV power density spectra of both the Suzaku and first XMM-Newton observations displayed a potential QPO component near 5 mHz. The fractional root-mean-square variability in the QPO is $\geq 2.8\%$ and $\sim 4\%$ for Suzaku and XMM-Newton, respectively. These two apparent quasi-periodic features are the motivation of our two-component jet model.

By studying the X-ray timing and spectral variability of Sw J1644+57,
\citet{sswk12} found that the X-ray spectrum became
much softer during each of the early dip, while the spectrum outside the dips became mildly
harder in its long-term evolution. Combining \citet{sswk12} with LZG13, we think that these soft dips
can be also interpreted by a spine-sheath two-component jet structure: a natural mechanism to account for these
dips would be a precessing jet driven by the B-P process at the Bardeen-Petterson radius $R_\mathrm{BP}$,
which is much softer than the inner jet launched at the BH horizon. The reason could be that the
Lorentz factor of the outer jet is much less than that of the inner jet,
leading to much softer photons during the early phase of each dip.

\begin{figure}[ht]
\centering
\includegraphics[width=80mm]{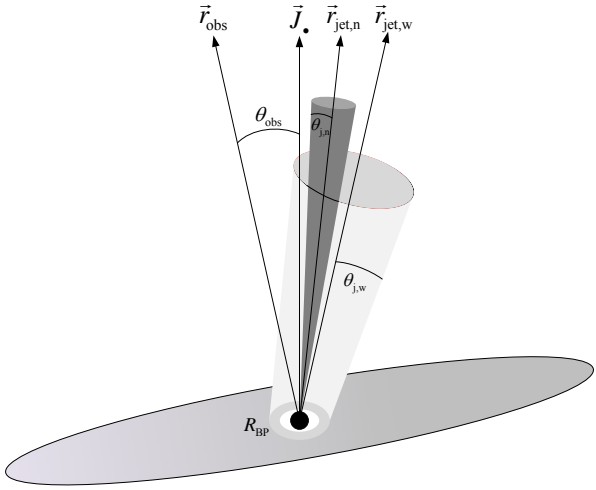}
\caption{A schematic drawing of the magnetic field configuration for a two-component jet, consisting of an inner
jet launched from the BH horizon and an outer jet launched from a warped accretion disk. The directions of both jets ($\vec{r}_\mathrm{jet,n}$ and $\vec{r}_\mathrm{jet,w}$ for the inner and outer jets, respectively): the outer jet is directed along
the disk normal direction around $R_\mathrm{BP}$, while the inner jet is directed along the direction of the magnetic axis of the BH, both are misaligned with the direction of the BH spin $\vec{J}_\bullet$.  The two jets are collimated with opening angles $\theta_\mathrm{j, n}$ and $\theta_\mathrm{j, w}$, respectively. The observer makes an angle $\theta_\mathrm{obs}$ with respect to the BH spin axis.}
\label{fig1}
\end{figure}

A two-component jet consists of an inner and an outer jet cone is shown in Fig.\ref{fig1}. The jet configuration is constructed base on following considerations:

(i) The initial stellar orbit is likely misaligned with the equatorial plane of the spinning BH.

(ii) The outer jet directed to the disk normal direction around $R_\mathrm{BP}$ as argued in LZG13. It is misaligned with the BH spin axis due to the Bardeen-Petterson effect (\citealt{bp75}; \citealt{wyy13}).

(iii) The inner jet is directed by the direction of the magnetic axis of BH. The BH alone can not maintain such magnetic field lines. They should be dragged by the accreting materials in the warped disk and are accumulated to the BH horizon. Therefore, the magnetic axis and then the inner jet is also misaligned (with a smaller degree with respect to the outer jet) with the direction of the BH spin $\vec{J}_\bullet$.

The inner jet is dragged by the spinning BH with an angular velocity $\Omega_\mathrm{F}$ of the magnetic field
lines, which is less than the angular velocity of the spinning BH $\Omega_\bullet$,
i.e., $\Omega_\mathrm{F}=k \Omega_\bullet$, with $k<1$.
Generally, the coefficient $k$ is uncertain, being taken as 0.5 for the optimal B-Z power (\citealt{mt82}; \citealt{lwm05}).
The period of the inner jet revolving around the BH spin can be estimated as
\begin{equation}\label{eq1}
  \tau_\mathrm{F}=2\pi / \Omega_\mathrm{F}=2\pi / (k \Omega_\bullet),
\end{equation}
the angular velocity of the BH is given by
\begin{equation}\label{eq2}
  \Omega_\bullet=a_\bullet / (2r_\bullet)=\frac{a_\bullet}{2(G M_\bullet/c^3)(1+\sqrt{1-a_\bullet^2})},
\end{equation}
where the BH spin $a_\bullet$ is related to the BH angular momentum by $J_\bullet=a_\bullet G M_\bullet^2/c$.

On the other hand, the precession period $\tau_\mathrm{P}$ of the outer jet can be estimated as
\begin{equation}\label{eq3}
  \tau_\mathrm{P}=2\pi / \Omega_\mathrm{LT},
\end{equation}
where $\Omega_\mathrm{LT}$ is the precession angular velocity due to the Lense-Thirring effect (hereafter LT, \citealt{lt18}),
and it reads
\begin{equation}\label{eq4}
  \Omega_\mathrm{LT}(R)=\frac{2G}{c^2}\frac{J_\bullet}{R^3}
  =2a_\bullet(\frac{c}{R})(\frac{R_\mathrm{g}}{R})^2.
\end{equation}

Taking the redshift factor $z = 0.354$ into account, we have the observational values of these two
periods as follows
\begin{equation}\label{eq5}
  \tau_\mathrm{F,obs}=(1+z)\tau_\mathrm{F}\simeq 200 (\rm s),
\end{equation}
\begin{equation}\label{eq6}
  \tau_\mathrm{P,obs}=(1+z)\tau_\mathrm{P}\simeq 2.7 (\rm days),
\end{equation}

It is easy to check that the ratio $\tau_\mathrm{P} / \tau_\mathrm{F}$ can be written as
\begin{equation}\label{eq7}
  \tau_\mathrm{P} / \tau_\mathrm{F}=\frac{k \xi^3}{4(1+\sqrt{1-a_\bullet^2})}\simeq 1.17 \times 10^3,
\end{equation}
where three parameters, $a_\bullet, k$ and $\xi$ are involved, and $\xi \equiv R_\mathrm{BP}/ R_\mathrm{g}$
is the radius $R_\mathrm{BP}$ in terms of the gravitational radius $R_\mathrm{g} \equiv GM_\bullet / c^2$.

Combining equations (\ref{eq1})--(\ref{eq7}) with $a_\bullet=0.9$ and $k =0.5$, we have the following set of the
parameters required by the two quasi-periodic variations (200s and 2.7-day) of Sw J1644+57:
\begin{equation}\label{eq8}
  M_\bullet \approx 7.5 \times 10^5 M_\odot , \;\;\;\;\;\;\; R=R_\mathrm{BP} \approx 23.8 R_\mathrm{g}.
\end{equation}

Fixing $a_\bullet=0.9$ and $R_\mathrm{BP} \approx 23.8 R_\mathrm{g}$, we have
$M_\bullet\approx 3.7 \times 10^5 M_\odot$ for $k=0.25$, and
$M_\bullet\approx 1.1 \times 10^6 M_\odot$ for $k=0.75$.
We therefore estimate (for $0.25 \leq k \leq 0.75$)
\begin{equation}\label{eq9}
  3.7 \times 10^5 M_\odot \leq M_\bullet \leq 1.1 \times 10^6 M_\odot, \;\;\;\;\;\;\; R=R_\mathrm{BP} \approx 23.8 R_\mathrm{g}.
\end{equation}

As argued in LZG13, the radius $R_\mathrm{BP}$ is closely related to the initial inclination of the stellar
orbit $\theta_\mathrm{orbit}$, and $R_\mathrm{BP} \approx 23.8 R_\mathrm{g}$ can be attained for appropriate values of $\theta_\mathrm{orbit}$ and the outer disk boundary radius $R_\mathrm{out}$.

Taking $k=0.5$ (corresponding to the optimal B-Z power), the estimated BH mass is
less than $10^6 M_\odot$ as shown in equation (\ref{eq8}). This estimation is roughly in accord with that given in
TMGK14 for the scenario of the partial tidal disruption of a low-mass main-sequence star,
but it is one order of magnitude smaller than that estimated in \citet{lz11}. However, the lower
BH mass does not lead to a lower jet luminosity, since $L_\mathrm{BZ}$ is essentially independent of the BH
mass as argued in \citet{lz11}.

\begin{figure}[ht]
\centering
\includegraphics[width=80mm]{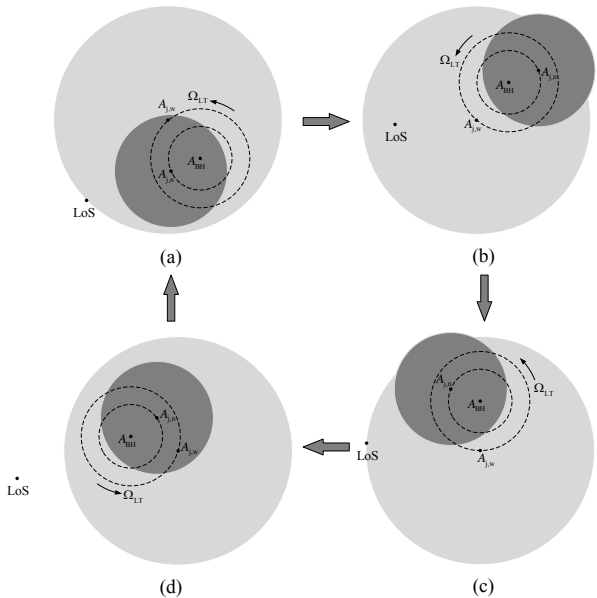}
\caption{The projected regions of the two jet cones (gray shaded circles) on the celestial sphere. Both jets revolve around the BH spin axis ($A_{\rm BH}$) along circular orbits (dashed circles). An observer is represented by a fixed line of sight (LoS). The outer jet cone (light gray shaded disc) revolves with an angular velocity $\Omega_{\rm LT}$ around $A_{\rm BH}$ along the larger dashed circle. The inner jet cone (dark disc) revolves with a much smaller period $\tau_{\rm F}$ around $A_{\rm BH}$ along the smaller dashed circle. The center of the outer and inner jets are denoted by $A_{\rm j,w}$ and $A_{\rm j,n}$, respectively.}
\label{fig2}
\end{figure}

In order to interpret the noticeable narrow dips in the 2.7-day quasi-periodic variation, we plot the projected regions of the two jet components on the celestial sphere, as shown in Fig.\ref{fig2}. The projected regions of the outer/inner jet cones are indicated by the larger/smaller shaded circles centered at $A_{\rm j,w}$/$A_{\rm j,n}$, respectively, and the BH spin axis $A_{\rm BH}$ and line of sight (LoS) are fixed points on the celestial sphere. It is found that the outer jet cone revolves with an angular velocity $\Omega_{\rm LT}$ around $A_{\rm BH}$ along the dashed circle given in Fig.\ref{fig2}, and the observer moves with a period $\tau_{\rm P}$, and enters and exits the outer jet cone periodically. The observer is located at inside the outer jet cone during the outer jet cone moves from panel (a) to panel (c), and is located at outside the outer jet cone during the outer jet cone leaves panel (c), passing by panel (d), and returns to panel (a). The narrow dips indicate that the LoS cannot be too far away from the BH spin axis $A_{\rm BH}$. One can expect $\theta_{\rm obs} < \theta_{\rm j, w}$ (LZG13). On the other hand,  the observer may remain outside the inner jet cone (as shown in Fig.\ref{fig2}), giving rise to much smaller variation in luminosity with a much smaller period $\tau_{\rm F}$ due to the revolution of inner jet around the BH spin. We thus take $\theta_\mathrm{j, n}<\theta_\mathrm{obs}<\theta_\mathrm{j, w}$. The main features of the two quasi-periodic variations of Sw J1644+57 can be then interpreted naturally by the two-component jet model.

A comparison between this and other models is summarized in Table \ref{tb1}. The advantage of this model lies in the fact that both quasi-periodic variations (200s QPO and 2.7-day variation) of Sw J1644+57 can be accounted for with a two-component jet model with a warped accretion disk, in which two promising jet launching mechanisms, the B-Z and B-P processes, are invoked.

\begin{table*}[htp]
\caption{Models for quasi-periodic variations of Swift J1644+57 \label{tb1}}
\begin{tabular}{ccccc}
\hline\noalign{\smallskip}
\hline\noalign{\smallskip}
    Models & Mechanisms & BH mass & Spin & Fitting \\
\hline\noalign{\smallskip}
    AL12 & Disk Resonance   & $\sim 10^5$ & $-1 < a_\bullet < 1$ & 200s QPO \\
    TMGK14 & MAD$^a$ + JD-QPO$^b$ & $10^5 \sim 10^6$ & $a_\bullet \geq 0.5$ & 200s QPO \\
    LZG13    & Precessing jet at warped disk & $2 \times 10^6$ & 0.9 & (2.7-day) variation \\
    This model    & B-Z Jet at BH, Precessing jet at warped disk      & $(0.37\sim 1.1)\times10^6$  & 0.9  & 200s QPO + (2.7-day) variation \\
\noalign{\smallskip}\hline
\end{tabular}

$^a$MAD is an abbreviation for ``magnetically arrested disk'' (\citealt{nia03})\\
$^b$JD-QPO is an abbreviation for “jet-disk quasi-periodic oscillation” (\citealt{mtb12})
\end{table*}

\section{Two-Component Jet Model for the Radio Light Curves and Spectra}

Sw J1644+57 was also accompanied by bright radio emission. The radio observations extending to $t \simeq 26$ days were first presented in \citet{zbs11}. Later \citet{bzp12} presented the data extending to $t \simeq 216$ days, and \citet{zbm13} presented the data extending to $t \simeq 600$ days.

The millimeter flux on a timescale of $\sim 100$ days is significantly brighter than what is expected by extrapolating the early declining light curve based on a single jet model (\citealt{mgm12, bzp12}). \citet{bzp12} and \citet{zbm13} suggested that this implies continuous injection of energy to the decelerating blast wave, which produced this re-brightening radiation for $\sim 100$ days. This model requires about 20 times more energy in the blast wave than the initial jet energy that produced the strong X-ray signal, which is of the order of $10^{52} \textrm{erg}$. It is puzzling why such a large kinetic energy did not leave imprints in X-rays.

As shown in \citet{lpl12}, this re-brightening can be naturally accounted for by a two-component jet model. However, their model is quite different from ours. They suggested that the inner jet is responsible for both the early X-ray emission and late radio re-brightening, while the outer jet is the source of the early times radio emission. In our model, as described above, both the inner and outer jets contribute to the X-ray emission, which leave imprints on the X-ray light curve in the form of quasi-periodic variations. The inner jet should be less energetic and is always outside of the line of sight as shown in Fig.\ref{fig2}. As a result, it is unlikely for it to account for the late re-brightening in radio emission. We attribute the late radio re-brightening to the more energetic outer jet.  \citet{lpl12} could only explain the radio flux up to 216 days. We will show that our model can interpret the long-term radio data up to $\sim 600$ days.

Inspecting the observed X-ray light curve of Sw J1644+57, the jet power may evolve with time as $\propto t^{-5/3}$ at later times. In such case, the late time dynamics of jet just depends on the total ejected kinetic isotropic-equivalent energy $E_\mathrm{k,iso}$, which is about the energy injected in the initial emission episode (\citealt{zm01}).

\begin{figure}[ht]
\centering
\includegraphics[width=80mm]{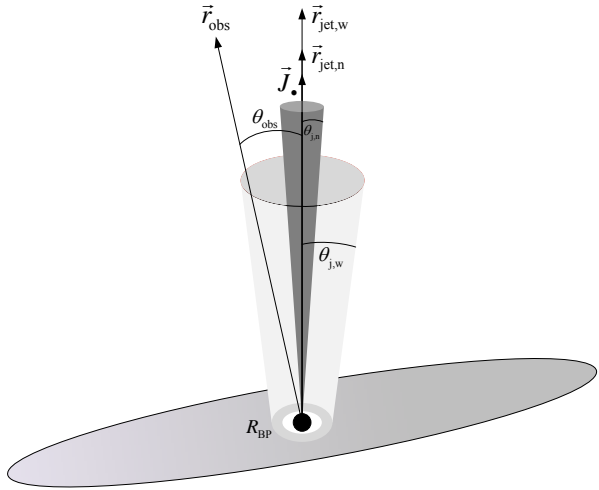}
\caption{A simpler version of Fig. 1, which is adopted to model the radio data. The directions of
the both jets are defined by the direction of the BH spin $\vec{J}_\bullet$. In this plot, $\theta_\mathrm{j, n}<\theta_\mathrm{obs}<\theta_\mathrm{j, w}$ is taken according to the discussion in Sect. 3.}
\label{fig3}
\end{figure}

For simplicity in the following calculations, we adopt a simplified jet geometry (as shown in Fig.\ref{fig3}). The two jets are assumed to be both along the BH spin axis. Comparing with Fig.\ref{fig1}, we now neglect the jet precession effect. Such a simple configuration is reasonable due to the follow reasons. First, the effect of jet precession does not noticeably change the jet dynamics, since we observe most of the outer-wide jet beam and are always outside the inner-narrow jet. Second, as discussed in \citet{tmgk14}, the precessing jet would evolve and finally be aligned with the BH spin axis. Finally, the observations also show the evolution of the precessing period (\citealt{sswk12, sm13}), which is consistent with that expected by \citet{tmgk14}.

We consider the two-component jet propagrates into a gaseous circumnuclear medium (CNM) with a number density $n$.
For a collimated jet, the jet effect becomes important when $1/\Gamma > \theta_\mathrm{j}$ \citep{zm04}. Studies of Sw J1644+57 indicated that the jet bulk Lorentz factor is $\Gamma \simeq 10$ and the jet opening angle is $\theta_\mathrm{j} \simeq 6^\circ$ (\citealt{bkg11, lzg13}). Therefore, we expect that the initial Lorentz factor obeys $\Gamma \theta_\mathrm{j} \le 1$. In the calculations, we include a suppression of the flux density by a factor of $(\Gamma \theta_\mathrm{j})^2 /2$ due to this jet break effect (\citealt{zm04}).

The steep spectral slope at lower frequencies indicate the existence of self-absorption (\citealt{zbs11, bzp12}). We therefore calculate the self-absorption frequency in detail.
\citet{kbbp13} argued that the inverse-Compton cooling of electrons by X-ray photons streaming through the external shock region is an important cooling mechanism. We therefore also consider this effect in our model. To do this, one needs to solve the following equation for the cooling Lorentz factor $\gamma_{\rm c}$,
\begin{equation}
\gamma_{\rm c} = \frac{3 m_{\rm e} c}{4\sigma_T U^\prime_{\rm B} (1+Y(\gamma_{\rm c})) \Gamma t/(1+z)},
\end{equation}
where
\begin{equation}
Y(\gamma_{\rm c}) = \frac{U'_{\rm ph,syn}+U'_{\rm ph,X}}{U^\prime_{\rm B}},
\end{equation}
$U^\prime_{\rm B} $ is comoving frame magnetic energy density, $U'_{\rm ph,syn}$ is the co-moving synchrotron photon energy density, and
\begin{equation}
U'_{\rm ph,X} = \frac{f^\prime_{\rm X} (<\nu'_{\rm kn})}{c} = \frac{L_{\rm X}(<\nu_{\rm kn})}{16\pi \Gamma^2 R^2 c}
\end{equation}
is X-ray photon energy density below the Klein-Nishina frequency, $\nu^\prime_{\rm kn} = m_{\rm e} c^2/(\gamma_{\rm c} h)$ in the co-moving frame, and $\nu_{\rm kn} = \Gamma \nu'_{\rm kn} / (1+z)$ in the observer frame. The spectral index in the X-ray band for Sw J1644+57 is $\beta \simeq 0.7$. Following \citet{kbbp13}, the X-ray luminosity in 0.3 -10 keV is taken as $L_{\rm X} = 4\times 10^{46} {\rm erg \ s^{-1} } (t/20 {\rm d})^{-5/3}$ for $t>20 $d, and as a constant between 10 and 20 d.

We also assume $p>2$. Due to poor baryon loading (\citealt{lzl13}), the inner jet should contain a large Lorentz factor. It is thus reasonable to assume that the early radio emission comes from the forward shock of the inner jet. The outer jet has more baryon loading and is more energetic, so that it decelerates at a later epoch, which would be responsible for the late rebrightening around $\sim 100$ days.

\begin{figure*}[htp]
\center
\includegraphics[width=160mm]{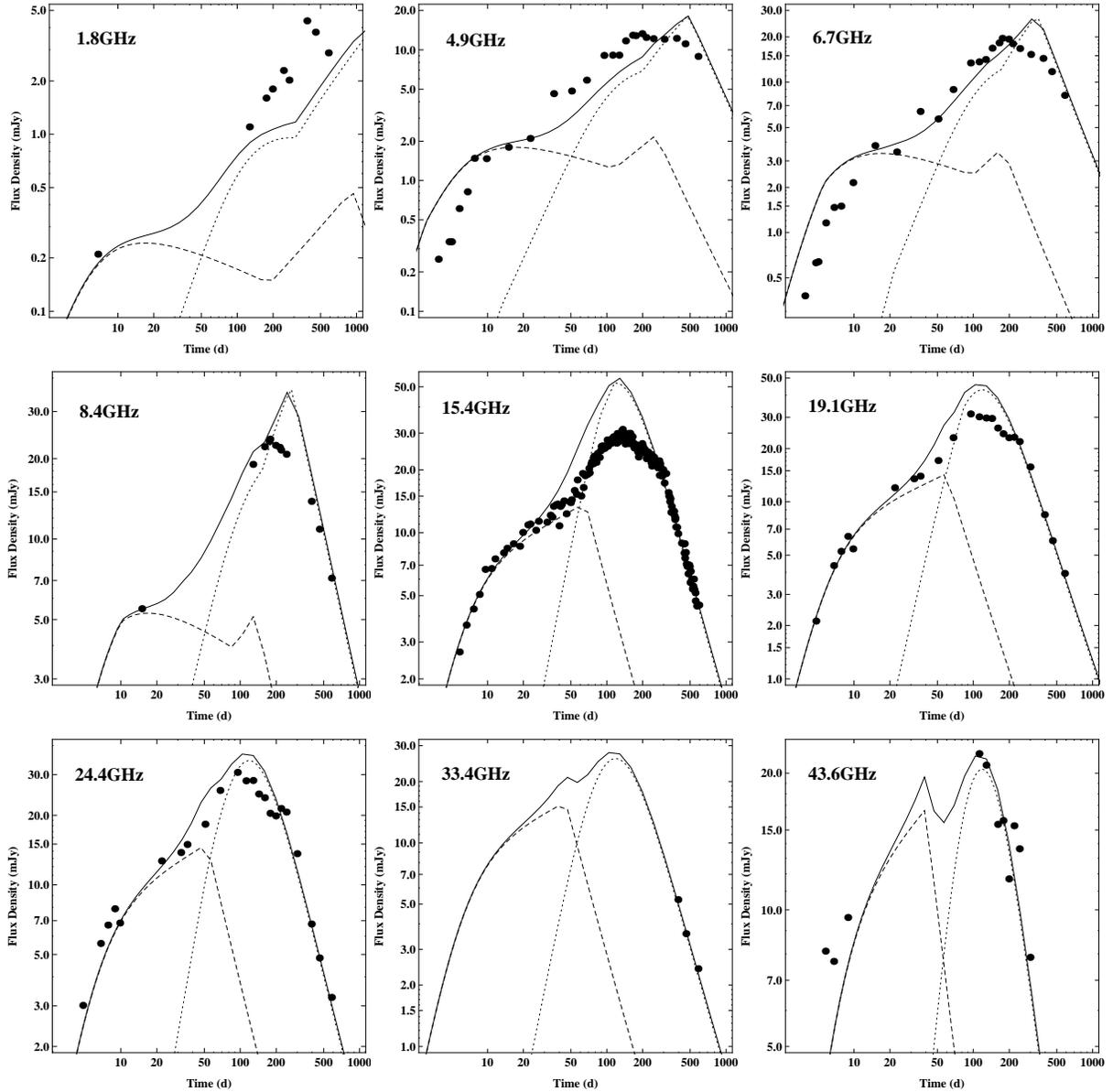}
\caption[]{Radio lightcurves of Sw J1644+57 extending to $t \simeq 600$ days. Observed data are from \citet{zbm13}. Each panel corresponds to a certain observed frequency as shown in the figure. The solid lines are the model lightcurves based on independent fits to the broad band SEDs (next figure) using the two-component jet model described in Section 3. The inner (outer) jet has kinetic isotropic-equivalent energy $E_{\rm k,iso} = $ $3.0 \times 10^{52}{\rm ergs}$ ($3.0 \times 10^{53}{\rm ergs}$), initial Lorentz factor $\Gamma_j=$ $5.5$ ($2.5$), opening angle $\theta_j=$ $6^{\circ}$ ($10^{\circ}$). The details of model parameters are shown in Table 2. The earlier peak is mainly contributed by the faster inner jet (dashed lines), while the later peak is mainly contributed by the slower outer jet (dotted lines).}
\label{fig_lc}
\end{figure*}

\begin{figure*}[htp]
\center
\includegraphics[width=150mm]{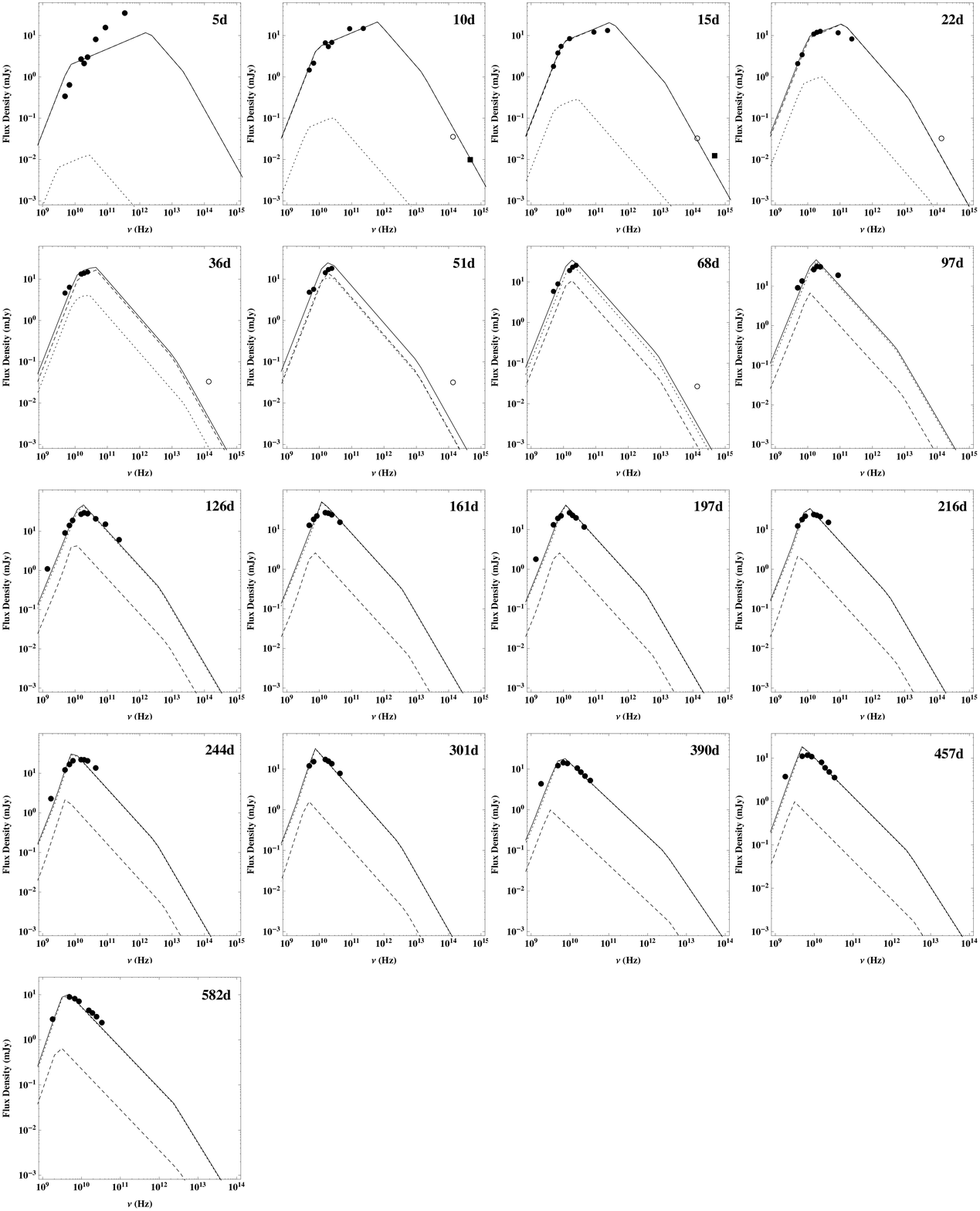}
\caption[]{ Multi-frequency radio (filled circles) spectral distributions of Sw J1644+57 at $t \simeq 5- 582$ days. The K/K$_s$ and R band flux at $t\simeq 10-68$ days are also shown with open circles and filled squares, respectively. The radio data are from \citet{zbm13}, and IR/optical data are from \citet{bkg11} and \citet{ltc11}. Each panel corresponds to a certain episode as shown in the figure. The solid lines are model fits based on our two-component jet model with the same parameters used in Figure \ref{fig_lc}. The contributions from the inner and outer jets are shown by the dashed and dotted lines, respectively.}
\label{fig_spec}
\end{figure*}

We develop a numerical two-component model to fit the data. The dynamical evolution of the jets are followed using a set of hydrodynamical equations \citep{hgdl00}. Synchrotron spectra of both jets are calculated using the standard broken-power-law spectral model (see below for more detailed discussion). We apply the model to simulately fit the light curves at observed frequencies of 1.8, 4.9, 6.7, 8.4, 15.4, 19.1, 24.4, 33.4 and 43.4 GHz, as presented in Fig.\ref{fig_lc}, and the broad-band spectra at 17 epochs during a span of $t=5 - 582$ days, as presented in Fig.\ref{fig_spec}. The fitting results indicate that the inner jet is narrower, less energetic but faster than the outer jet. The best-fit parameters are the following. For the inner jet, the isotropic-equivalent kinetic energy is $E_{\rm k,iso} = 3.0\times 10^{52}{\rm erg}$, the initial Lorentz factor is $\Gamma_j = 5.5$, and the jet opening angle $\theta_\mathrm{j}=6^{\circ}$; for the outer jet, one has $E_{\rm k,iso} = 3.0\times 10^{53}{\rm erg}$, $\Gamma_j = 2.5$, and $\theta_\mathrm{j}=6^{\circ}$.
The CNM density is $n = 0.25 \, {\rm cm^{-3}}$, and the observer's viewing angle is $\theta_{\rm obs}=7^{\circ}$. Table 2 summarize the model parameters of the two jet components.

\begin{table*}[htp]
\begin{center}
\caption{Results of radio observations fits \label{tb2}}
\begin{tabular}{ccccccccc}
\hline\noalign{\smallskip}
\hline\noalign{\smallskip}
    Jet component & n (${\rm cm}^{-3}$) & $\theta_{\rm obs}$ (deg) & $E_{52}$ & $\Gamma_{\rm j}$ & $\theta_{\rm j}$ (deg) & $p$ & $\epsilon_{\rm B}$ & $\epsilon_{\rm e}$ \\
\hline\noalign{\smallskip}
    inner-narrow & 0.25 & 7.0 &3.0   & 5.5 & 6.0 & 2.8 & 0.25 & 0.2 \\
    outer-wide & 0.25 & 7.0 &30.0   & 2.5 & 10.0 & 2.8 & 0.13 & 0.15 \\
\noalign{\smallskip}\hline
\end{tabular}
\end{center}
\end{table*}

\begin{figure}[ht]
\centering
\includegraphics[width=80mm]{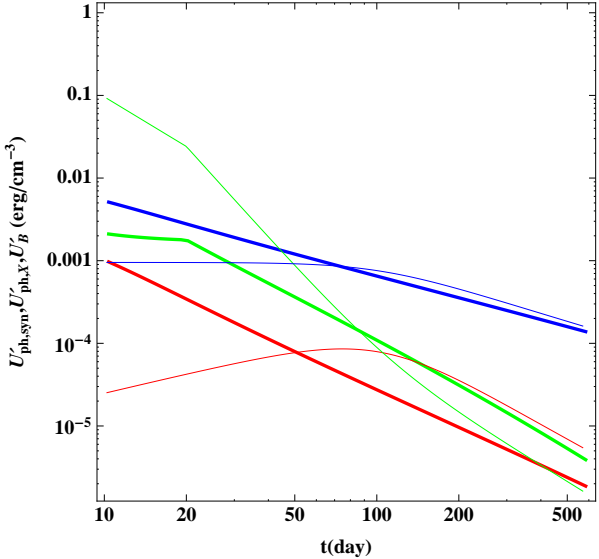}
\caption{The time evolution of the co-moving frame magnetic field energy density $U'_{\rm B}$ (blue lines), synchrotron photon energy density $U'_{\rm ph,syn}$ (red lines), and X-ray photon energy density $U'_{\rm ph,X}$ (green lines). Thick and thin lines are the results for the inner and outer jet, respectively. It is found that the IC cooling is not important  at $t>10$ days for the inner jet and $t>50$ days for the outer jet. For both jets,  SSC effects can be neglected.}
\label{fig:U}
\end{figure}

As argued by \citet{kbbp13}, the spectra and lightcurve would be significantly modified if the IC cooling of electrons by X-ray photons becomes important. In our modeling, we fully take into account the cooling effect due to synchrotron, SSC, and this ``external'' inverse Compton (EIC) process. With our best fit model, it turns out that both EIC and SSC effects are not important. This is because our best-fit model invokes large $\epsilon_B$ vaues in both jets (0.25 in the inner jet and 0.13 in the outer jet), so that the magnetic field energy density dominates over the photon energy densities in the emission region during most of time.  For comparison, \cite{kbbp13} adopted a much lower $\epsilon_{\rm B}$ ($4.5 \times 10^{-3}$) so that the EIC cooling effect is significant. In Fig.\ref{fig:U}, we compare the X-ray photon energy density $U^\prime_{\rm ph,X} = f^\prime_{\rm X}/c$, synchrotron photon energy density $U'_{\rm ph,syn}$ with the magnetic energy density $U^\prime_{\rm B}$, which shows that the EIC cooling effect is not important at $t>10$ days for the inner jet and $t>50$ days for the outer jet, and SSC can be ignored for both jets. In our model, the outer jet becomes dominating only at late times when EIC cooling can be neglected. However, the early suppression of $\nu_{\rm c}$ by EIC cooling improves the fit to the infrared observations. For example, the flux in the infrared K-band at 10 days changes from $0.07 \ \rm mJy$ to $0.04 \ \rm mJy$ after considering the EIC cooling, which is more consistent with the observed K-band flux of $0.03 \ \rm mJy$ at this time (\citealt{ltc11}). As shown in Fig.\ref{fig_spec}, our model predictions do not exceed the observed optical flux, which likely has a contribution from the host light (\citealt{bkg11,ltc11}).

In the following, we provide an analytical analysis of the problem, which catches the main picture of jet evolution. For simplicity, we do not include IC cooling in the scaling laws (see above). Our equations generally follow the recent review article of Gao et al. (2013b) on analytical synchrotron radiation models of GRBs.

Both jets first undergo a coasting phase, in which we have $\Gamma (t) = \Gamma_\mathrm{j}$, and $R(t) = 2c \Gamma_\mathrm{j}^2 t$. Based on the evolution for $\Gamma (t)$, we can give expressions for the time evolution of synchrotron frequencies as (\citealt{wdhl03, glzwz13}),
\begin{eqnarray}
&&\nu_{\rm m}=   1.3 \times10^{13}~{\rm Hz}~\Gamma_{\rm j,1}^{4} \epsilon_{e,-1}^{2}\epsilon_{\rm B,-1}^{1/2}           ,\nonumber\\
&&\nu_{\rm c}=    2.0 \times10^{14}~{\rm Hz}~\Gamma_{\rm j,1}^{-4}  \epsilon_{B,-1}^{-3/2} t_{d}^{-2}           \nonumber\\
&&\nu_{\rm a} =8.7 \times10^{9}~{\rm Hz}~ \Gamma_{\rm j,1}^{8/5} \epsilon_{e,-1}^{-1}\epsilon_{B,-1}^{1/5}t_{d}^{3/5}, ~~~~\nu_a < \nu_m < \nu_c\nonumber\\
\label{eq:coasting}
\end{eqnarray}
where $z=0.354$, $p = 2.8$, and $n=0.25 \rm cm^{-3}$ are used. The above equations can be generally applied to both the inner and outer jets. For the inner one, however, the frequencies (times) should be divided (multiplied) by a factor $f_{\rm view}= (1-\beta)/(1-\beta \cos\theta_{\rm v})$, which takes into account the correction factor of an off-axis jet with respect to an on-axis one. Here $\beta=\sqrt{1-1/\Gamma^2}$ is the dimensionless velocity of the inner jet, and $\theta_{\rm v}$ is the angle between the near-edge of the inner jet and the observer. For the parameters given in Table 2, $f_{\rm view}$ is found to be around 1.1 .

According to Eq. (\ref{eq:coasting}), the lower radio frequency emission would undergo a transition from optically thin to optically thick during this phase. The synchrotron flux is given by:
\begin{eqnarray}
&& F_\nu = 2.8 ~ {\rm mJy}~ \Gamma_{j,1}^{26/3} \epsilon_{e,-1}^{-2/3}\epsilon_{B,-1}^{1/3} \theta_{j,-1}^2 \nu_{10}^{1/3} t_d^3 , ~ \nu_a<\nu<\nu_m \nonumber\\
&& F_\nu = 3.6 ~ {\rm mJy}~ \Gamma_{j,1}^{6} \epsilon_{e,-1} \theta_{j,-1}^2 \nu_{10}^{2} t_d^2
, ~~~~~~~~~~~~
\nu<\nu_a<\nu_m \nonumber\\
\end{eqnarray}
Noticing the sharp dependence on $\Gamma$ in these regimes, one can immediately draw the conclusion that the emission in this phase is dominated by the inner jet, which has a larger Lorentz factor.

The jets start to decelerate when the mass collected from the CNM is about $1/\Gamma$ of the rest mass entrained in the ejecta. The deceleration time of the ejecta with an isotropic kinetic energy $E_{\rm k,iso}$ and an initial Lorentz factor $\Gamma_{\rm j}$ is
\begin{equation}
t_{\rm dec} = (1+z) \left[ \frac{3E_{\rm k,iso} }{16 \pi n m_{\rm p} \Gamma_{\rm j}^8 c^5 } \right]^{1/3} \simeq   1.3 \ {\rm day} \  E_{52}^{1/3}  \Gamma_{\rm j, 1}^{-8/3},
\end{equation}
where $E_{52}$ donates $E_{\rm k,iso,52}$ for simplicity. This corresponds to about 10 days for the inner jet and 200 days for the outer jet. After this time, the two jets approach the \citet{bm76} self-similar evolution\footnote{Note that BM solution was adopted just for the analytical presentation of our results. In reality this phase is not fully developed since the jets are mildly relativistic. Our numerical model has fully incorporated the transition from relativistic to non-relativistic phases, which was used to fit the lightcurves and SEDs.},
\begin{eqnarray}
&& \Gamma (t) \simeq 4.9  E_{ 52}^{1/8}  t_d^{-3/8}, \nonumber\\
&& R (t) \simeq 5.1 \times 10^{17}  \ {\rm cm} \  E_{ 52}^{1/4}  t_d^{1/4}.
\label{BM}
\end{eqnarray}
At this stage, the characteristic synchrotron frequencies are give by (\citealt{glzwz13}):
\begin{eqnarray}
&&\nu_m=   7.1\times10^{11}~{\rm Hz}~E_{52}^{1/2}\epsilon_{e,-1}^{2}\epsilon_{B,-1}^{1/2} t_{d}^{-3/2}           ,\nonumber\\
&&\nu_c=    3.5\times10^{15}~{\rm Hz}~E_{52}^{-1/2} \epsilon_{B,-1}^{-3/2}t_{d}^{-1/2},           \nonumber\\
&&\nu_a=9.6\times10^{9}~{\rm Hz}~E_{52}^{1/5} \epsilon_{e,-1}^{-1} \epsilon_{B,-1}^{1/5}, ~~~~~~~~~~
\nu_a < \nu_m < \nu_c\nonumber\\
&&\nu_a=9.3 \times10^{10}~{\rm Hz}~ E_{52}^{0.35} \epsilon_{e,-1}^
{0.53}\epsilon_{B,-1}^{0.35} t_{d}^{-0.76}, ~\nu_m < \nu_a < \nu_c . \nonumber \\
\end{eqnarray}
One can see that $\nu_m$ decreases very quickly, so that the jets would evolve from the regime $\nu_a<\nu_m<\nu_c$ to $\nu_m<\nu_a<\nu_c$ after the deceleration.

During the first a few tens of days, the inner jet decelerates while the outer jet is still in the pre-deceleration stage. The emission is dominated by synchrotron radiation from the inner jet.  Since it occurs at earlier times, we have the flux from the inner jet as,
\begin{equation}
F_\nu = 2.8  ~{\rm mJy}~ E_{52}^{13/12} \epsilon_{e,-1}^
{-2/3} \epsilon_{B,-1}^{1/3}  \theta_{j,-1}^2 \nu_{10}^{1/3} t_d^{-1/4},
\end{equation}
for $\nu_a<\nu<\nu_m<\nu_c$. At $\sim 43$ days, the emission from the outer jet becomes the dominant component. The light curve rises first (during the coasting phase), and then decays after the jet enters the deceleration phase at about 200 days. The late-time flux of the outer jet is given by (in the regime of $\nu_m<\nu_a<\nu_c$):
\begin{eqnarray}
&& F_\nu = 27114 ~{\rm mJy}~ E_{53}^{1.7} \epsilon_{e,-1}^
{1.8} \epsilon_{B,-1}^{0.95} \theta_{j,-1}^2 \nu_{10}^{-0.9} t_d^{-2.1} , ~ \nu_a<\nu<\nu_c \nonumber\\
&& F_\nu = 0.88 ~{\rm mJy}~ E_{53}^{0.5} \epsilon_{B,-1}^{-0.25} \theta_{j,-1}^2 \nu_{10}^{5/2} t_d^{0.5}
, ~~~~~~~ \nu_m<\nu<\nu_a \nonumber\\
\end{eqnarray}

The blastwave eventually enters the Newtonian phase when it has swept up a CNM mass whose rest energy is comparable to the energy of the ejecta. The Sedov time is,
\begin{equation}
t_{\rm Sedov} = (1+z) \frac{3}{17} \left[ \frac{3E_{\rm k,iso} }{4 \pi n m_{\rm p}  c^5 } \right]^{1/3} \simeq 368 \ {\rm day} \   E_{53}^{1/3} ,
\end{equation}
For the inner (outer) jet, the Sedov time is about 200 days (400 days), which is much longer than that estimated by Eq.(\ref{BM}) (e.g., taken to be the time when $\Gamma \le 2$, i.e., $\sim 11 \ {\rm day} \   E_{52}^{1/3} $). This is reasonable, because the BM solution is not fully applicable for such a mildly relativistic case.

In the non-relativistic (Newtonian) regime, the dynamics is described by the well know Sedov-Taylor solution. We have the synchrotron frequencies as (\citealt{glzwz13})
\begin{eqnarray}
&&\nu_m=   1.3\times10^{17}~{\rm Hz}~E_{53} \epsilon_{e,-1}^{2}\epsilon_{B,-1}^{1/2}t_{d}^{-3}           ,\nonumber\\
&&\nu_c=    1.6\times10^{14}~{\rm Hz}~E_{53}^{-3/5}\epsilon_{B,-1}^{-3/2}t_{d}^{-1/5} ,          \nonumber\\
&&\nu_a= 1.1 \times10^{12}~{\rm Hz}~E_{53}^{0.4} \epsilon_{e,-1}^{0.53}\epsilon_{B,-1}^{0.35} t_{d}^{-0.94},
~~\nu_m < \nu_a < \nu_c\nonumber \\
\end{eqnarray}
In this phase, the synchrotron emission is dominated by the outer jet.  The synchrotron emission flux from the outer jet can be written as
\begin{eqnarray}
&& F_\nu = 95712 ~{\rm mJy}~ E_{53}^{1.7} \epsilon_{e,-1}^{1.8}  \epsilon_{B,-1}^{0.95} \theta_{j,-1}^2 \nu_{10}^{-0.9} t_d^{-2.1} , ~ \nu_a<\nu<\nu_c \nonumber\\
&& F_\nu = 0.01 ~{\rm mJy}~ E_{53}^{0.3}  \epsilon_{B,-1}^{-0.25} \theta_{j,-1}^2 \nu_{10}^{5/2} t_d^{1.1}
, ~~~~~~~ \nu_m<\nu<\nu_a \nonumber\\
\end{eqnarray}

This two-component external shock model also predicts X-ray emission. As a self-consistency check, we also compare the predicted X-ray flux with the long-term X-ray observational data. Late time X-ray observations with Swift/XRT and Chandra revealed a dramatic change in the lightcurve evolution: following a steady $t^{-5/3}$ decline at $t \simeq 15-500$ days, there was a sharp decline by about a factor of 170 at $t>500-610$ days.

\begin{figure}[ht]
\centering
\includegraphics[width=80mm]{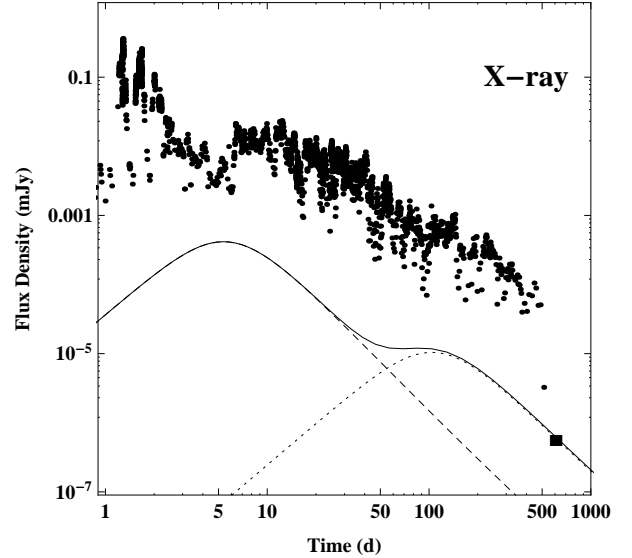}
\caption{X-ray lightcurve from Swift/XRT (filled circles) and a late-time Chandra observation (filled square). A rapid decline in the X-ray flux is evident at $t > 500$ days. The solid line shows the X-ray emission expected from the forward shock using the synchrotron model described in Section 3. The dashed and dotted lines are the contributions from the inner and outer jets, respectively. The model indicates that the flux measured in the Chandra observation is consistent with the external shock emission from the outer jet.}
\label{fig:xray}
\end{figure}

In Fig.\ref{fig:xray}, we plot the X-ray emission expected from the forward shock and compare it with the observed X-ray lightcurve with both Swift/XRT and Chandra. The observed X-ray emission at $t<500$ days is much brighter than the model prediction, and exhibits a rapid variability. This suggest that it is not from the external shock, but likely arises from the internal dissipation of the jet (\citealt{bkg11, bgm11,zbs11,mgm12,lpl12,zwc13}). This is consistent with our arguments in Sect.2 to interpret the quasi-periodic variations of X-ray emission. The rapid decline at $t<500$ days also disfavors the external shock origin of early X-ray emission.
One may therefore conclude that the central engine has been shut off at $t \sim 500$ days (\citealt{zbm13}).

It is interesting to note that the last Chandra point is somewhat higher than the simple extrapolation of the steep decay phase, but is in excellent agreement with the external forward shock model prediction. From Figure \ref{fig:xray}, we can see that this fit would fail without the wide jet. This result lends further support to our two-component jet model.

\section{Conclusions and Discussion}
In this paper, we study the two quasi-periodic variations (200s and 2.7-day) in the X-ray emission and the rebrightening in late-time radio observations of Sw J1644+57.  The two quasi-periodic variations with very different periods led us to speculate a two-component jet model. Such a model is also motivated by two natural jet-launching mechanisms, i.e., the B-Z process for powering the jet from the BH horizon and the B-P process for driving an outflow from the warped disk at $R_\mathrm{BP}$
(\citealt{bp82,lzg13}).
This two-component jet scenario naturally predicts two afterglow components, which nicely interpret the radio data of the source, especially the mysterious re-brightening $\sim 100$ days after the trigger.

The significant re-brightening in radio emission at $t>100$ days put strong constrains on the single jet model. \citet{bzp12} and \citet{zbm13} introduced an energy injection model to interpret this excess. Alternatively, \citet{kbbp13} suggested that the radio emitting electrons suffer inverse Compton cooling by the X-ray photons passing through the blastwave region. Within this scenario, the increase of the effective $\epsilon_{\rm e}$ as the X-ray flux decreases with time (so that the cooling mechanism weakens) is responsible for the apparent energy increase in the radio emitting region. Considering that the jet may be magnetically dominated (\citealt{bkg11, szfw11, lz11, gzz12}), \citet{dp13} suggested a narrow Poynting flux dominated jet model in which its magnetic energy gradually converts to particle energy (e.g. \citealt{zy11}). They argued that this model may solve the energy budget problem faced in many other models. Although these models may explain the radio data, the two-component jet model proposed here is a natural one, without the need of invoking a huge energy injection at late times or introducing unconventional ingredients in the afterglow models. Compared with the two-component jet model proposed by \citet{lpl12}, our model can interpret both the two quasi-periodic variations in X-rays and the entire radio observations (including data points later than 200 days).

Since the inner (B-Z) and outer (B-P) jets move relatively with each other through precession and differential motion due to different Lorentz factors, dissipations of magnetic energy in both jets are inevitable. One can speculate two types of interactions. One is ``collision'' between the inner jet and the outer jet as the former streams into the later, the other is the relative ``shearing'' at the boundary of two jets. These would induce significant magnetic dissipation, through processes similar to the ICMART process envisaged in GRBs (e.g. \citealt{zy11}) or something similar to Kelvin-Helmholtz instability (in the strongly magnetized regime). It is difficult to quatitatively calculate the strength of such dissipations. On the other hand, these dissipation processes provide a natural mechanism to power the observed X-ray emission in the jet. One may estimate the dissipation efficiency from data, using the X-ray observations and the radio data fits. The total X-ray energy emitted from Sw J1644+57 is $E_{\rm X, iso} \simeq 2\times 10^{53}$ erg in the 1.35-13.5 keV band. Based on our fits to the radio data of Sw J1644+57, the total kinetic energy of the outer jet is $E_{\rm K, iso} = 3\times 10^{53}$ erg. The total jet energy can be therefore estimated as $E_{\rm j, iso} = E_{\rm X, iso}+E_{\rm K, iso}   \simeq 5\times 10^{53}$ erg. Therefore, the magnetic dissipation efficiency could be estimated as $\eta =  E_{\rm X, iso} /E_{\rm j, iso} \simeq 0.4$.  Such an efficiency is quite high, comparable to that of GRBs (e.g. \citealt{zlp07}), and consistent with the magnetic jet dissipation models \citep[e.g.][]{zy11}.

Even though our model can fit most of the data, from Fig.\ref{fig_spec} one can see that it fails to explain the observations around 5 days. For the light curves, we can not reproduce the sharp peak at $\sim 500$ days in the 1.8 GHz data. For 4.9, 6.7, 8.4 and 15.4 GHz lightcurves, our model curves exhibit too sharp peaks.
These discrepancies may arise from the simplified forward shock model we adopted here.

In most previous papers, a wind like medium was adopted in the modeling (\citealt{zbs11, mgm12, bzp12, zbm13, kbbp13}). Such a stratified medium profile was introduced to better fit the early-time radio data (e.g., Metzger et al. 2012). Some other authors adopted a constant medium density profile (e.g. Liu et al. 2012; Zou et al. 2013). In this work, we try to fit the entire radio data set, with focus more on the late-time radio observations. We find that in general a constant medium density profile provides an overall better fit than the wind-like stratified profile.  It is possible that the medium is stratified nearby the central BH, probably modified by a galactic wind, but becomes a constant density one at larger radii when the influence of the wind becomes negligible.

\citet{kbbp13} suggested that the inverse-Compton cooling off the X-ray photons has a significant effect on electrons in the external shock region, which may result in flat light curves in the radio and mm bands. Depending parameters, synchrotron self-Compton \citep{glwz13} can be also important.
On the other hand, we found that both EIC and SSC processes do not play an important role in our modeling. This is because our best fit requires a large $\epsilon_{\rm B}$. We have compared the magnetic field energy density with the photon densities in the emission region, and found that in the temporal regimes relevant to the radio observations, the IC processes can be neglected. However, the early suppression of $\nu_c$ by EIC cooling improves the fit to the infrared observations.

Besides the B-P mechanism for launching a jet from the disk, it is possible that an intrinsically episodic jet may be launched from the disk through a magnetic process (\citealt{yz12}). This can also produce variable light curves in X-rays. We do not discuss this mechanism in this paper.

The rich observations of Sw J1644+57 allow us to study this TDE source with relativistic jet in great detail. Another candidate Sw J2058+05 shared many similarities with Sw J1644+57 (Cenko et al. 2012), which deserves further investigation. There may be also some TDEs hidden in the GRB samples (e.g. Lu et al. 2008; Gao et al. 2010; Levan et al. 2014). Detailed observations and theoretical modeling are needed to identify these sources.

\acknowledgements  This work is supported by National Basic Research Program (``973'' Program) of China under Grant Nos. 2014CB845800 and 2009CB824800, National Natural Science Foundation of China under grants 11361140349 (China-Israel jointed program), 11003004, 11173011 and U1231101, NSF under Grant No. AST-0908362, and Fundamental Research Funds for the Central Universities (HUST: 2013QN021). WHL acknowledges an Open Research Program from the Key Laboratory for the Structure and Evolution of Celestial Objects of Chinese Academy of Sciences for support.



\clearpage

\clearpage

\end{document}